\documentclass[10pt]{article}
\usepackage{graphicx}
\usepackage{amsmath}
\usepackage{amssymb}
\usepackage{caption2}
\begin{document}
\bibliographystyle{prsty}
\begin{center}
{\large {\bf \sc{  Analysis of  the hidden-charm pentaquark candidates in the $J/\psi p$ mass spectrum  with QCD sum rules }}} \\[2mm]
Zhi-Gang Wang \footnote{E-mail: zgwang@aliyun.com.  }     \\
 Department of Physics, North China Electric Power University, Baoding 071003, P. R. China
\end{center}

\begin{abstract}
 In this work, we distinguish the isospin unambiguously to construct the   diquark-diquark-antiquark type five-quark currents with the isospin $I=\frac{1}{2}$,  and study the $uudc\bar{c}$ pentaquark states with the QCD sum rules systematically for the first time. Then we obtain the mass spectrum of the  diquark-diquark-antiquark type $uudc\bar{c}$ pentaquark states with the isospin-spin-parity $IJ^P=\frac{1}{2}{\frac{1}{2}}^-$, $\frac{1}{2}{\frac{3}{2}}^-$ and $\frac{1}{2}{\frac{5}{2}}^-$, and make  possible assignments of the $P_c(4312)$, $P_c(4337)$, $P_c(4380)$, $P_c(4440)$ and $P_c(4457)$ states.   As a byproduct, we obtain the lowest hidden-charm pentaquark state which lies just above the $\bar{D}\Lambda_c$ threshold.
\end{abstract}



\section{Introduction}
In 2015,  the  LHCb collaboration  observed  two pentaquark candidates $P_c(4380)$ and $P_c(4450)$ in the $J/\psi p$ invariant mass distribution in the $\Lambda_b^0\to J/\psi K^- p$ decays with  significances of more than $9\,\sigma$  \cite{LHCb-4380}. The  measured Breit-Wigner masses and widths are,
\begin{flalign}
 &P_c(4380) : M = 4380\pm 8\pm 29 \mbox{ MeV}\, , \, \Gamma = 205\pm 18\pm 86 \mbox{ MeV} \, , \nonumber \\
  &P_c(4450) : M = 4449.8\pm 1.7\pm 2.5 \mbox{ MeV} \, ,\, \Gamma = 39\pm 5\pm 19 \mbox{ MeV} \,   .
\end{flalign}
They  have the preferred spin-parity   $J^P={\frac{3}{2}}^-$ and ${\frac{5}{2}}^+$, respectively, however,  the assignments of the  spin-parity     $J^P={\frac{3}{2}}^+$ and ${\frac{5}{2}}^-$ cannot be excluded.

In 2019, the LHCb collaboration studied the $\Lambda_b^0\to J/\psi K^- p$ decays with a data sample, which is an order of magnitude larger than the previously one, and observed a new pentaquark candidate $P_c(4312)$  with a significance of  $7.3\sigma$ \cite{LHCb-Pc4312}. Furthermore,
 the LHCb collaboration confirmed the $P_c(4450)$ pentaquark structure, which  consists  of two narrow overlapping peaks $P_c(4440)$ and $P_c(4457)$
  with  a significance of  $5.4\sigma$ \cite{LHCb-Pc4312}.
   The measured Breit-Wigner masses and widths are
\begin{flalign}
 &P_c(4312) : M = 4311.9\pm0.7^{+6.8}_{-0.6} \mbox{ MeV}\, , \, \Gamma = 9.8\pm2.7^{+ 3.7}_{- 4.5} \mbox{ MeV} \, , \nonumber \\
 & P_c(4440) : M = 4440.3\pm1.3^{+4.1}_{-4.7} \mbox{ MeV}\, , \, \Gamma = 20.6\pm4.9_{-10.1}^{+ 8.7} \mbox{ MeV} \, , \nonumber \\
 &P_c(4457) : M = 4457.3\pm0.6^{+4.1}_{-1.7} \mbox{ MeV} \, ,\, \Gamma = 6.4\pm2.0_{- 1.9}^{+ 5.7} \mbox{ MeV} \,   .
\end{flalign}
The spin and parity are not determined.

In 2021, the LHCb collaboration observed evidences  for a new structure in the $J/\psi p$ and $J/\psi \bar{p}$ invariant mass distributions  with a significance in the range of 3.1 to 3.7$\sigma$,  which depend on the assigned $J^P$ hypothesis \cite{LHCb-Pc4337}, the measured Breit-Wigner mass and width are
\begin{flalign}
  &P_c(4337) : M = 4337 \ ^{+7}_{-4} \ ^{+2}_{-2} \mbox{ MeV} \, ,\, \Gamma = 29 \ ^{+26}_{-12} \ ^{+14}_{-14} \mbox{ MeV} \,   .
\end{flalign}

The $P_c(4312)$, $P_c(4337)$, $P_c(4380)$, $P_c(4440)$ and $P_c(4457)$ were observed in the $J/\psi p$ invariant mass distributions, if the strong interactions conserve the isospin exactly, they should have the isospin $I=\frac{1}{2}$, while the $P_{cs}(4338)$ and $P_{cs}(4459)$ observed in the $J/\psi\Lambda$ invariant mass distributions  should have the isospin $I=0$ \cite{LHCb-Pcs4338,LHCb-Pcs4459-2012}.

The $P_c(4312)$,  $P_c(4380)$, $P_c(4440)$ and $P_c(4457)$ states lie at the $\bar{D}\Sigma_c$, $\bar{D}\Sigma_c^*$, $\bar{D}^*\Sigma_c$ and $\bar{D}^*\Sigma_c$ thresholds, respectively, while the $P_{cs}(4338)$ and $P_{cs}(4459)$ lie at the
$\bar{D}\Xi_c$ and $\bar{D}^*\Xi_c$ thresholds, respectively. We naively expect that they are the hadronic molecular states \cite{mole-penta-1,mole-penta-2,mole-penta-3,mole-penta-4,Pc4312-mole-penta-1,
    mole-penta-5,Pc4312-mole-penta-2,Pc4312-mole-penta-3,Pc4312-mole-penta-4,
    Pc4312-mole-penta-5,Pc4312-mole-penta-6,mole-penta-6,Pc4312-mole-penta-7,
mole-penta-8,mole-penta-9,mole-penta-10,mole-penta-11,Pc4312-mole-penta-8,Pc4312-mole-penta-WXW-SCPMA,
Pc4312-mole-penta-WXW-IJMPA,Pcs4338-mole-FKGuo}.   As the $P_c(4337)$ is concerned,   it lies not far way from  the  $\bar{D}^* \Lambda_c$,
  $\bar{D} \Sigma_c$ and $\bar{D} \Sigma_c^*$ thresholds, but it does not lie just in any baryon-meson  threshold, it is very difficult to assign it as a molecular state without introducing large coupled channel effects. The molecule scenario still needs fine-tuning to work, the diquark-diquark-antiquark type pentaquark scenario \cite{di-di-anti-penta-1,Wang1508-EPJC,WZG-penta-IJMPA,Pc4312-penta-1,di-di-anti-penta-2,
WangHuang-EPJC-1508-12,WangZG-EPJC-1509-12,WangZG-NPB-1512-32,WangZhang-APPB,di-di-anti-penta-3,
Pc4312-penta-3} and  diquark-triquark  type  pentaquark scenario \cite{di-tri-penta-1,di-tri-penta-2} are all robust candidates, other interpretations such as anomalous triangle singularities are also feasible \cite{ATS-LiuXH-2016,ATS-Bayar-GuoFK-2016,ATS-Mikhasenko-2015}.

In the photoproduction, those resonances might  be produced
in the s-channel $\gamma p \to P_c \to J/\psi p$ with the vector meson dominance, which is expected to be free from the three-body re-scattering effects \cite{PhotonPr-Karliner-2016,PhotonPr-Hiller-2016,PhotonPr-XYWang-2019}, but
there might be cusp structures at the $\Lambda_c \bar{D}^{(*)}$ thresholds in the energy dependence of the $J/\psi$ photoproduction cross section due to the intermediate states $\Lambda_c \bar{D}^{(*)}$ \cite{PhotonPr-cusp-FKGuo-2020}.
Experimentally, the GlueX collaboration observed no evidence for those pentaquark states in recent $J/\psi$-photoproduction \cite{PhotonPr-GlueX-PRL-2019,PhotonPr-GlueX-PRC-2023}. The peak structures $P_c(4312)$, $P_c(4380)$ and $P_c(4457)$ could be reproduced via the double triangle amplitudes
through an interference with other common mechanisms \cite{PhotonPr-GlueX-Explain-2021}, it is a possible interpretation of the GlueX experiment. In the off-shell
coupled-channel formalism, Clymton, Kim and Mart obtained the transition amplitudes for the $J/\psi
N$ scattering and various heavy meson and singly heavy baryon
scattering processes, and they observed that those $P_c$ states
undergo significant modifications in the $J/\psi N$ elastic
channel, with some even disappearing due to interference from the
positive parity channel \cite{PhotonPr-GlueX-HCKim-2024}, which could provide an
insight into the absence of the $P_c$ states in the GlueX experiments. And such a mechanism was extended to study the $P_{cs}$ states \cite{PhotonPr-HCKim-2025}.

We expect to obtain a suitable and uniform scheme to accommodate  all the existing pentaquark candidates. In the present work, we would like to focus on the diquark-diquark-antiquark type pentaquark scenario and resort to the QCD sum rules.

The QCD sum rules approach is a powerful theoretical tool in exploring the exotic states, such as the tetraquark states, pentaquark states, molecular states, etc \cite{Nielsen-review,WangZG-Review}.
  In Refs.\cite{Wang1508-EPJC,WangHuang-EPJC-1508-12,WangZG-EPJC-1509-12,WangZG-NPB-1512-32,WangZhang-APPB}, we study the diquark-diquark-antiquark type hidden-charm pentaquark states with  the spin-parity $J^P={\frac{1}{2}}^\pm$, ${\frac{3}{2}}^\pm$, ${\frac{5}{2}}^\pm$   and the strangeness  $S=0,\,-1,\,-2,\,-3$ via the QCD sum rules in an systematic way, but do not exhaust all the possible configurations and a lot of works are still needed. We calculate the  vacuum condensates up to  dimension 10 and adopt the energy scale formula \cite{WangHuang3900,Wang-tetra-formula,WangZG-mole-formula-1,WangZG-mole-formula-2,
Wang-tetra-PRD-HC,Wang-tetra-NPB-HCss,WangZG-Landau-PRD},
  \begin{eqnarray}
\mu =\sqrt{M_{P}^2-(2{\mathbb{M}}_c)^2} \, ,
 \end{eqnarray}
 to choose the best energy scales of the QCD spectral densities, where the ${\mathbb{M}}_c$ is the effective $c$-quark mass.
  The resulting Borel platforms are not flat enough, as the higher dimensional vacuum condensates play a very  important role in acquiring the flat platforms.

After the discovery of the $P_c(4312)$,  we updated the old analysis by accomplishing  the operator product expansion up to the   vacuum condensates of $13$ consistently  and adopting the updated value of the effective $c$-quark mass ${\mathbb{M}}_c=1.82\,\rm{GeV}$ \cite{WZG-tetraquark-Mc},  and  restudied the ground state mass spectrum of the diquark-diquark-antiquark type $uudc\bar{c}$ pentaquark states with the QCD sum rules, and revisit the assignments of the $P_c$ states  \cite{WZG-penta-IJMPA}. However, we did not distinguish the isospin and studied  the five-quark configurations with the isospins $I=\frac{1}{2}$ and $\frac{3}{2}$ together, and did not exhaust  the lowest configurations with either the isospin $I=\frac{1}{2}$ or $\frac{3}{2}$.

After the discovery of the $P_{cs}(4459)$, we studied the possibility of assigning it as the isospin cousin of the $P_{c}(4312)$ by taking account of the light-flavor $SU(3)$ breaking effects \cite{WangZG-Pcs4459-333}, then we studied  the diquark-diquark-antiquark type $udsc\bar{c}$ with the isospin $I=0$ and spin-parity $J^P={\frac{1}{2}}^-$, ${\frac{3}{2}}^-$ and ${\frac{5}{2}}^-$ in a comprehensive way and try to assign the $P_{cs}(4338)$ and $P_{cs}(4459)$ in the scenario of diquark-diquark-antiquark type  pentaquark states consistently \cite{WangZG-XinQ-Pcs}.

In this work, we try to exhaust  the lowest   diquark-diquark-antiquark type $uudc\bar{c}$ pentaquark configurations with the isospin $I=\frac{1}{2}$, and study them with the QCD sum rules systematically, and revisit  the possible assignments of the $P_{c}$ states with the isospin $I=\frac{1}{2}$ and  spin-parity $J^P={\frac{1}{2}}^-$, ${\frac{3}{2}}^-$ or ${\frac{5}{2}}^-$.

The article is arranged as follows: we acquire  the QCD sum rules for the  pentaquark states with the isospin $I=\frac{1}{2}$ in Sect.2;  in Sect.3, we present the numerical results and discussions; and Sect.4 is reserved for our
conclusion.

\section{QCD sum rules for  the  $uudc\bar{c}$ pentaquark states}
Firstly, let us write down  the  correlation functions  $\Pi(p)$,  $\Pi_{\mu\nu}(p)$ and $\Pi_{\mu\nu\alpha\beta}(p)$,
\begin{eqnarray}\label{CF-Pi-Pi-Pi}
\Pi(p)&=&i\int d^4x e^{ip \cdot x} \langle0|T\left\{J(x)\bar{J}(0)\right\}|0\rangle \, ,\nonumber\\
\Pi_{\mu\nu}(p)&=&i\int d^4x e^{ip \cdot x} \langle0|T\left\{J_{\mu}(x)\bar{J}_{\nu}(0)\right\}|0\rangle \, ,\nonumber\\
\Pi_{\mu\nu\alpha\beta}(p)&=&i\int d^4x e^{ip \cdot x} \langle0|T\left\{J_{\mu\nu}(x)\bar{J}_{\alpha\beta}(0)\right\}|0\rangle \, ,
\end{eqnarray}
where $J(x)=J^1(x)$, $J^2(x)$, $J^3(x)$, $J^4(x)$, $J_\mu(x)=J_\mu^1(x)$, $J_\mu^2(x)$, $J_\mu^3(x)$, $J_\mu^4(x)$, $J_{\mu\nu}(x)=J_{\mu\nu}^1(x)$, $J_{\mu\nu}^2(x)$,
\begin{eqnarray}
 J^1(x)&=&\varepsilon^{ila} \varepsilon^{ijk}\varepsilon^{lmn}  u^T_j(x) C\gamma_5 d_k(x)\,u^T_m(x) C\gamma_5 c_n(x)\,  C\bar{c}^{T}_{a}(x) \, , \nonumber\\
J^2(x)&=&\varepsilon^{ila} \varepsilon^{ijk}\varepsilon^{lmn}  u^T_j(x) C\gamma_5 d_k(x)\,u^T_m(x) C\gamma_\mu c_n(x)\,\gamma_5 \gamma^\mu C\bar{c}^{T}_{a}(x) \, ,\nonumber\\
J^{3}(x)&=&\frac{\varepsilon^{ila} \varepsilon^{ijk}\varepsilon^{lmn}}{\sqrt{2}} \left[u^T_j(x) C\gamma_\mu u_k(x)d^T_m(x) C\gamma^\mu c_n(x) -u^T_j(x) C\gamma_\mu d_k(x)u^T_m(x) C\gamma^\mu c_n(x) \right] C\bar{c}^{T}_{a}(x) \, , \nonumber\\
J^{4}(x)&=&\frac{\varepsilon^{ila} \varepsilon^{ijk}\varepsilon^{lmn}}{\sqrt{2}} \left[u^T_j(x) C\gamma_\mu u_k(x) d^T_m(x) C\gamma_5 c_n(x)-u^T_j(x) C\gamma_\mu d_k(x) u^T_m(x) C\gamma_5 c_n(x)\right] \gamma_5 \gamma^\mu  C\bar{c}^{T}_{a}(x) \, ,\nonumber\\
\end{eqnarray}
for the isospin-spin $(I,J)=(\frac{1}{2},\frac{1}{2})$,
\begin{eqnarray}
J^1_\mu(x)&=&\varepsilon^{ila} \varepsilon^{ijk}\varepsilon^{lmn}  u^T_j(x) C\gamma_5 d_k(x)\,u^T_m(x) C\gamma_\mu c_n(x) C\bar{c}^{T}_{a}(x) \, ,\nonumber\\
J^{2}_{\mu}(x)&=&\frac{\varepsilon^{ila} \varepsilon^{ijk}\varepsilon^{lmn}}{\sqrt{2}} \left[ u^T_j(x) C\gamma_\mu u_k(x) d^T_m(x) C\gamma_5 c_n(x)-u^T_j(x) C\gamma_\mu d_k(x) u^T_m(x) C\gamma_5 c_n(x)\right]    C\bar{c}^{T}_{a}(x) \, ,  \nonumber\\
 J^{3}_{\mu}(x)&=&\frac{\varepsilon^{ila} \varepsilon^{ijk}\varepsilon^{lmn}}{\sqrt{2}} \left[u^T_j(x) C\gamma_\mu u_k(x)d^T_m(x) C\gamma_\alpha c_n(x) -u^T_j(x) C\gamma_\mu d_k(x)u^T_m(x) C\gamma_\alpha c_n(x) \right] \gamma_5\gamma^\alpha C\bar{c}^{T}_{a}(x) \, ,\nonumber\\
J^{4}_{\mu}(x)&=&\frac{\varepsilon^{ila} \varepsilon^{ijk}\varepsilon^{lmn}}{\sqrt{2}} \left[ u^T_j(x) C\gamma_\alpha u_k(x)d^T_m(x) C\gamma_\mu c_n(x) -u^T_j(x) C\gamma_\alpha d_k(x)u^T_m(x) C\gamma_\mu c_n(x) \right] \gamma_5\gamma^\alpha C\bar{c}^{T}_{a}(x) \, , \nonumber\\
\end{eqnarray}
for the isospin-spin $(I,J)=(\frac{1}{2},\frac{3}{2})$,
\begin{eqnarray}
J^1_{\mu\nu}(x)&=&\frac{\varepsilon^{ila} \varepsilon^{ijk}\varepsilon^{lmn} }{\sqrt{2}} u^T_j(x) C\gamma_5 d_k(x)\left[u^T_m(x) C\gamma_\mu c_n(x)\, \gamma_5\gamma_{\nu}C\bar{c}^{T}_{a}(x)+u^T_m(x) C\gamma_\nu c_n(x)\,\gamma_5 \gamma_{\mu}C\bar{c}^{T}_{a}(x)\right] \, ,\nonumber\\
J^2_{\mu\nu}(x)&=&\frac{\varepsilon^{ila} \varepsilon^{ijk}\varepsilon^{lmn}}{2} \left[ u^T_j(x) C\gamma_\mu u_k(x)d^T_m(x) C\gamma_\nu c_n(x)-u^T_j(x) C\gamma_\mu d_k(x)u^T_m(x) C\gamma_\nu c_n(x) \right]C\bar{c}^{T}_{a}(x)    \nonumber\\
 &&+\left( \mu\leftrightarrow\nu\right)\, ,  \nonumber\\
\end{eqnarray}
for the isospin-spin $(I,J)=(\frac{1}{2},\frac{5}{2})$,
where the $i$, $j$, $k$, $l$, $m$, $n$ and $a$ are color indexes, the $C$ is the charge conjugation matrix. We take the currents $J^1(x)$, $J^2(x)$, $J^1_{\mu}(x)$ and $J^1_{\mu\nu}(x)$ from Ref.\cite{WZG-penta-IJMPA} and update the analysis, and construct  other five-quark currents with the isospin $I=\frac{1}{2}$ to study the possible hidden-charm pentaquark mass spectrum  in the $J/\psi p$ invariant  mass distribution, as the strong decays conserve isospin in general. There exists a $u$-$d$ quark pair in each current, which is anti-symmetric  under the interchange $u\leftrightarrow d$ and warrants the zero isospin, the residue $u$-quark provides an isospin $I=\frac{1}{2}$, thus the currents have the total isospin $I=\frac{1}{2}$.

In the currents $J(x)$, $J_\mu(x)$ and $J_{\mu\nu}(x)$, there are diquark constituents  $\varepsilon^{ijk}u^T_jC\gamma_5d_k$, $\varepsilon^{ijk}q^T_jC\gamma_{\mu}q_k$, $\varepsilon^{ijk}q^T_jC\gamma_5c_k$, $\varepsilon^{ijk}q^T_jC\gamma_{\mu}c_k$ with $q=u$ or $d$, the most stable diquark configurations, therefore we could obtain the lowest hidden-charm pentaquark configurations.
 The light diquarks  $\varepsilon^{ijk}u^T_jC\gamma_5d_k$ and $\varepsilon^{ijk}q^T_jC\gamma_{\mu}q_k$ have the spins $S_L=0$ and $1$, respectively,  the heavy diquarks $\varepsilon^{ijk}q^T_jC\gamma_5c_k$ and $\varepsilon^{ijk}q^T_jC\gamma_{\mu}c_k$ have the spins $S_H=0$ and $1$, respectively. The light and  heavy diquarks  form a charmed tetraquark in color triplet with the  angular momentum $\vec{J}_{LH}=\vec{S}_L+\vec{S}_H$, and  $J_{LH}=0$, $1$ or $2$.
The $\bar{c}$-quark operators $C\bar{c}_a^T$ and $\gamma_5\gamma_{\mu}C\bar{c}_a^T$ have the spin-parity $J^P={\frac{1}{2}}^-$ and ${\frac{3}{2}}^-$, respectively. The total angular momentums $\vec{J}=\vec{J}_{LH}+\vec{J}_{\bar{c}}$ with the values $J=\frac{1}{2}$, $\frac{3}{2}$ or $\frac{5}{2}$, see Table \ref{current-pentaQ}.
In fact, there exists another current $J^5_\mu(x)$,
\begin{eqnarray}\label{Jmu-5}
J^5_\mu(x)&=&\varepsilon^{ila} \varepsilon^{ijk}\varepsilon^{lmn}  u^T_j(x) C\gamma_5 d_k(x)\,u^T_m(x) C\gamma_5 c_n(x)\, \gamma_5\gamma_{\mu} C\bar{c}^{T}_{a}(x) \, ,
\end{eqnarray}
which leads to the same QCD sum rules as the current $J^1(x)$ up to a numerical factor, the $[ud][uc]\bar{c}$ ($0$, $0$, $0$, $\frac{1}{2}$) state with the spin-parity $J^P={\frac{1}{2}}^-$ and $[ud][uc]\bar{c}$ ($0$, $0$, $0$, $\frac{3}{2}$) state with the spin-parity $J^P={\frac{3}{2}}^-$ might have degenerated masses.

\begin{table}
\begin{center}
\begin{tabular}{|c|c|c|c|c|c|c|c|c|}\hline\hline
$[qq][qc]\bar{c}$ ($S_L$, $S_H$, $J_{LH}$, $J$)& $J^{P}$            & Currents              \\ \hline

$[ud][uc]\bar{c}$ ($0$, $0$, $0$, $\frac{1}{2}$)                     & ${\frac{1}{2}}^{-}$  & $J^1(x)$              \\

$[ud][uc]\bar{c}$ ($0$, $1$, $1$, $\frac{1}{2}$)                     & ${\frac{1}{2}}^{-}$  & $J^2(x)$              \\

$[uu][dc]\bar{c}-[ud][uc]\bar{c}$ ($1$, $1$, $0$, $\frac{1}{2}$)     & ${\frac{1}{2}}^{-}$  & $J^3(x)$              \\

$[uu][dc]\bar{c}-[ud][uc]\bar{c}$ ($1$, $0$, $1$, $\frac{1}{2}$)     & ${\frac{1}{2}}^{-}$  & $J^4(x)$              \\

$[ud][uc]\bar{c}$ ($0$, $1$, $1$, $\frac{3}{2}$)                     & ${\frac{3}{2}}^{-}$  & $J^1_\mu(x)$           \\

$[uu][dc]\bar{c}-[ud][uc]\bar{c}$ ($1$, $0$, $1$, $\frac{3}{2}$)     & ${\frac{3}{2}}^{-}$  & $J^2_\mu(x)$          \\

$[uu][dc]\bar{c}-[ud][uc]\bar{c}$ ($1$, $1$, $2$, $\frac{3}{2}$)${}_3$  & ${\frac{3}{2}}^{-}$  & $J^3_\mu(x)$           \\

$[uu][dc]\bar{c}-[ud][uc]\bar{c}$ ($1$, $1$, $2$, $\frac{3}{2}$)${}_4$  & ${\frac{3}{2}}^{-}$  & $J^4_\mu(x)$           \\

$[ud][uc]\bar{c}$ ($0$, $1$, $1$, $\frac{5}{2}$)                     & ${\frac{5}{2}}^{-}$  & $J^1_{\mu\nu}(x)$     \\

$[uu][dc]\bar{c}-[ud][uc]\bar{c}$ ($1$, $1$, $2$, $\frac{5}{2}$)     & ${\frac{5}{2}}^{-}$  & $J^2_{\mu\nu}(x)$      \\ \hline\hline
\end{tabular}
\end{center}
\caption{ The quark structures and spin-parity of the  currents.  }\label{current-pentaQ}
\end{table}

The currents $J(x)$, $J_\mu(x)$ and $J_{\mu\nu}(x)$ have negative parity, but they couple potentially to both the negative and positive parity hidden-charm pentaquark states with the isospin $I=\frac{1}{2}$, as   multiplying $i \gamma_{5}$ to the currents  $J(x)$, $J_\mu(x)$ and $J_{\mu\nu}(x)$ changes their parity \cite{WangZG-Review}.
Let us write down the current-hadron couplings explicitly,
\begin{eqnarray}\label{Coupling12}
\langle 0| J (0)|P_{\frac{1}{2}}^{-}(p)\rangle &=&\lambda^{-}_{\frac{1}{2}} U^{-}(p,s) \, , \nonumber \\
\langle 0| J (0)|P_{\frac{1}{2}}^{+}(p)\rangle &=&\lambda^{+}_{\frac{1}{2}} i\gamma_5 U^{+}(p,s) \, ,
\end{eqnarray}
\begin{eqnarray}
\langle 0| J_{\mu} (0)|P_{\frac{3}{2}}^{-}(p)\rangle &=&\lambda^{-}_{\frac{3}{2}} U^{-}_\mu(p,s) \, ,  \nonumber \\
\langle 0| J_{\mu} (0)|P_{\frac{3}{2}}^{+}(p)\rangle &=&\lambda^{+}_{\frac{3}{2}}i\gamma_5 U^{+}_\mu(p,s) \, ,  \nonumber \\
\langle 0| J_{\mu} (0)|P_{\frac{1}{2}}^{+}(p)\rangle &=&f^{+}_{\frac{1}{2}}p_\mu U^{+}(p,s) \, , \nonumber \\
\langle 0| J_{\mu} (0)|P_{\frac{1}{2}}^{-}(p)\rangle &=&f^{-}_{\frac{1}{2}}p_\mu i\gamma_5 U^{-}(p,s) \, ,
\end{eqnarray}
\begin{eqnarray}\label{Coupling52}
\langle 0| J_{\mu\nu} (0)|P_{\frac{5}{2}}^{-}(p)\rangle &=&\sqrt{2}\lambda^{-}_{\frac{5}{2}} U^{-}_{\mu\nu}(p,s) \, ,\nonumber\\
\langle 0| J_{\mu\nu} (0)|P_{\frac{5}{2}}^{+}(p)\rangle &=&\sqrt{2}\lambda^{+}_{\frac{5}{2}}i\gamma_5 U^{+}_{\mu\nu}(p,s) \, ,\nonumber\\
\langle 0| J_{\mu\nu} (0)|P_{\frac{3}{2}}^{+}(p)\rangle &=&f^{+}_{\frac{3}{2}} \left[p_\mu U^{+}_{\nu}(p,s)+p_\nu U^{+}_{\mu}(p,s)\right] \, , \nonumber\\
\langle 0| J_{\mu\nu} (0)|P_{\frac{3}{2}}^{-}(p)\rangle &=&f^{-}_{\frac{3}{2}}i\gamma_5 \left[p_\mu U^{-}_{\nu}(p,s)+p_\nu U^{-}_{\mu}(p,s)\right] \, , \nonumber\\
\langle 0| J_{\mu\nu} (0)|P_{\frac{1}{2}}^{-}(p)\rangle &=&g^{-}_{\frac{1}{2}}p_\mu p_\nu U^{-}(p,s) \, , \nonumber\\
\langle 0| J_{\mu\nu} (0)|P_{\frac{1}{2}}^{+}(p)\rangle &=&g^{+}_{\frac{1}{2}}p_\mu p_\nu i\gamma_5 U^{+}(p,s) \, ,
\end{eqnarray}
where the subscripts $\frac{1}{2}$, $\frac{3}{2}$ and $\frac{5}{2}$ denote the spins, the superscripts $\pm$ denote the positive and negative parity, respectively, the $\lambda$, $f$ and $g$ are the pole residues.
The spinors $U^\pm(p,s)$ satisfy the Dirac equations  $(\not\!\!p-M_{\pm})U^{\pm}(p)=0$, while the spinors $U^{\pm}_\mu(p,s)$ and $U^{\pm}_{\mu\nu}(p,s)$ satisfy the Rarita-Schwinger equations $(\not\!\!p-M_{\pm})U^{\pm}_\mu(p)=0$ and $(\not\!\!p-M_{\pm})U^{\pm}_{\mu\nu}(p)=0$ \cite{Wang1508-EPJC,WZG-penta-IJMPA,WangZG-Review,Wang-cc-baryon-penta}.

At the hadron  side, we insert  a complete set  of intermediate pentaquark states with the same quantum numbers as the interpolating currents  $J(x)$, $i\gamma_5 J(x)$,
$J_{\mu}(x)$, $i\gamma_5 J_{\mu}(x)$, $J_{\mu\nu}(x)$ and $i\gamma_5 J_{\mu\nu}(x)$  into the correlation functions
$\Pi(p)$, $\Pi_{\mu\nu}(p)$ and $\Pi_{\mu\nu\alpha\beta}(p)$ to obtain the hadronic representations
\cite{SVZ79-1,SVZ79-2,PRT85}, and  isolate the  lowest hidden-charm pentaquark states, and obtain the  results:
\begin{eqnarray}\label{CF-Hadron-12}
\Pi(p) & = & {\lambda^{-}_{\frac{1}{2}}}^2  {\!\not\!{p}+ M_{-} \over M_{-}^{2}-p^{2}  }+  {\lambda^{+}_{\frac{1}{2}}}^2  {\!\not\!{p}- M_{+} \over M_{+}^{2}-p^{2}  } +\cdots  \, ,\nonumber\\
&=&\Pi_{\frac{1}{2}}^1(p^2)\!\not\!{p}+\Pi_{\frac{1}{2}}^0(p^2)\, ,
 \end{eqnarray}
\begin{eqnarray}\label{CF-Hadron-32}
 \Pi_{\mu\nu}(p) & = & {\lambda^{-}_{\frac{3}{2}}}^2  {\!\not\!{p}+ M_{-} \over M_{-}^{2}-p^{2}  } \left(- g_{\mu\nu}+\cdots
\right)+  {\lambda^{+}_{\frac{3}{2}}}^2  {\!\not\!{p}- M_{+} \over M_{+}^{2}-p^{2}  } \left(- g_{\mu\nu} +\cdots
\right)   \nonumber \\
& &+ {f^{+}_{\frac{1}{2}}}^2  {\!\not\!{p}+ M_{+} \over M_{+}^{2}-p^{2}  } p_\mu p_\nu+  {f^{-}_{\frac{1}{2}}}^2  {\!\not\!{p}- M_{-} \over M_{-}^{2}-p^{2}  } p_\mu p_\nu  +\cdots  \, , \nonumber\\
&=&\left[\Pi_{\frac{3}{2}}^1(p^2)\!\not\!{p}+\Pi_{\frac{3}{2}}^0(p^2)\right]\left(- g_{\mu\nu}\right)+\cdots\, ,
\end{eqnarray}
\begin{eqnarray}\label{CF-Hadron-52}
\Pi_{\mu\nu\alpha\beta}(p) & = &2{\lambda^{-}_{\frac{5}{2}}}^2  {\!\not\!{p}+ M_{-} \over M_{-}^{2}-p^{2}  } \left[\frac{ \widetilde{g}_{\mu\alpha}\widetilde{g}_{\nu\beta}+\widetilde{g}_{\mu\beta}\widetilde{g}_{\nu\alpha}}{2}
+\cdots\right]+  2 {\lambda^{+}_{\frac{5}{2}}}^2  {\!\not\!{p}- M_{+} \over M_{+}^{2}-p^{2}  } \left[\frac{ \widetilde{g}_{\mu\alpha}\widetilde{g}_{\nu\beta}+\widetilde{g}_{\mu\beta}\widetilde{g}_{\nu\alpha}}{2}
+\cdots \right]\nonumber\\
&&+  {f^{+}_{\frac{3}{2}}}^2  {\!\not\!{p}+ M_{+} \over M_{+}^{2}-p^{2}  } \left[ p_\mu p_\alpha \left(- g_{\nu\beta}+\cdots\right)+\cdots \right]+  {f^{-}_{\frac{3}{2}}}^2  {\!\not\!{p}- M_{-} \over M_{-}^{2}-p^{2}  } \left[ p_\mu p_\alpha \left(- g_{\nu\beta}+\cdots\right)+\cdots \right]   \nonumber \\
& &+ {g^{-}_{\frac{1}{2}}}^2  {\!\not\!{p}+ M_{-} \over M_{-}^{2}-p^{2}  } p_\mu p_\nu p_\alpha p_\beta+  {g^{+}_{\frac{1}{2}}}^2  {\!\not\!{p}- M_{+} \over M_{+}^{2}-p^{2}  } p_\mu p_\nu p_\alpha p_\beta  +\cdots \, , \nonumber\\
& = & \left[\Pi_{\frac{5}{2}}^1(p^2)\!\not\!{p}+\Pi_{\frac{5}{2}}^0(p^2)\right]\left( g_{\mu\alpha}g_{\nu\beta}+g_{\mu\beta}g_{\nu\alpha}\right)  +\cdots \, ,
 \end{eqnarray}
where $\widetilde{g}_{\mu\nu}=g_{\mu\nu}-\frac{p_{\mu}p_{\nu}}{p^2}$. We study  the components $\Pi_{\frac{1}{2}}^1(p^2)$, $\Pi_{\frac{1}{2}}^0(p^2)$, $\Pi_{\frac{3}{2}}^1(p^2)$, $\Pi_{\frac{3}{2}}^0(p^2)$, $\Pi_{\frac{5}{2}}^1(p^2)$ and  $\Pi_{\frac{5}{2}}^0(p^2)$ to avoid possible contaminations from other pentaquark states with different spins.

We obtain the hadronic spectral densities  through  dispersion relation,
\begin{eqnarray}
\frac{{\rm Im}\Pi^1_j(s)}{\pi}&=& \lambda_{-}^2 \delta\left(s-M_{-}^2\right)+\lambda_{+}^2 \delta\left(s-M_{+}^2\right) =\, \rho^1_{H}(s) \, , \\
\frac{{\rm Im}\Pi^0_j(s)}{\pi}&=&M_{-}\lambda_{-}^2 \delta\left(s-M_{-}^2\right)-M_{+}\lambda_{+}^2 \delta\left(s-M_{+}^2\right)
=\rho^0_{H}(s) \, ,
\end{eqnarray}
where $j=\frac{1}{2}$, $\frac{3}{2}$, $\frac{5}{2}$, we introduce the subscript $H$ to denote   the hadron side,
then we introduce the  weight functions $\sqrt{s}\exp\left(-\frac{s}{T^2}\right)$ and $\exp\left(-\frac{s}{T^2}\right)$ to obtain the QCD sum rules
at the hadron side,
\begin{eqnarray}
\int_{4m_c^2}^{s_0}ds \left[\sqrt{s}\,\rho^1_{H}(s)+\rho^0_{H}(s)\right]\exp\left( -\frac{s}{T^2}\right)
&=&2M_{-}\lambda_{-}^2\exp\left( -\frac{M_{-}^2}{T^2}\right) \, ,
\end{eqnarray}
\begin{eqnarray}
\int_{4m_c^2}^{s^\prime_0}ds \left[\sqrt{s}\,\rho^1_{H}(s)-\rho^0_{H}(s)\right]\exp\left( -\frac{s}{T^2}\right)
&=&2M_{+}\lambda_{+}^2\exp\left( -\frac{M_{+}^2}{T^2}\right) \, ,
\end{eqnarray}
where the $s_0$ and $s_0^\prime$ are the continuum threshold parameters, and the $T^2$ is the Borel parameter. Thus we separate    the  contributions  of the hidden-charm pentaquark states with negative and positive parity unambiguously.

At the QCD side,  we accomplish  the operator product expansion with the help of the full $u$, $d$ and $c$ quark propagators,
 \begin{eqnarray}
U/D_{ij}(x)&=& \frac{i\delta_{ij}\!\not\!{x}}{ 2\pi^2x^4}-\frac{\delta_{ij}\langle
\bar{q}q\rangle}{12} -\frac{\delta_{ij}x^2\langle \bar{q}g_s\sigma Gq\rangle}{192} -\frac{ig_sG^{a}_{\alpha\beta}t^a_{ij}(\!\not\!{x}
\sigma^{\alpha\beta}+\sigma^{\alpha\beta} \!\not\!{x})}{32\pi^2x^2} -\frac{\delta_{ij}x^4\langle \bar{q}q \rangle\langle g_s^2 GG\rangle}{27648} \nonumber\\
&&  -\frac{1}{8}\langle\bar{q}_j\sigma^{\mu\nu}q_i \rangle \sigma_{\mu\nu}+\cdots \, ,
\end{eqnarray}
\begin{eqnarray}
C_{ij}(x)&=&\frac{i}{(2\pi)^4}\int d^4k e^{-ik \cdot x} \left\{
\frac{\delta_{ij}}{\!\not\!{k}-m_c}
-\frac{g_sG^n_{\alpha\beta}t^n_{ij}}{4}\frac{\sigma^{\alpha\beta}(\!\not\!{k}+m_c)+(\!\not\!{k}+m_c)
\sigma^{\alpha\beta}}{(k^2-m_c^2)^2}\right.\nonumber\\
&&\left. -\frac{g_s^2 (t^at^b)_{ij} G^a_{\alpha\beta}G^b_{\mu\nu}(f^{\alpha\beta\mu\nu}+f^{\alpha\mu\beta\nu}+f^{\alpha\mu\nu\beta}) }{4(k^2-m_c^2)^5}+\cdots\right\} \, ,\nonumber\\
f^{\alpha\beta\mu\nu}&=&(\!\not\!{k}+m_c)\gamma^\alpha(\!\not\!{k}+m_c)\gamma^\beta(\!\not\!{k}+m_c)\gamma^\mu(\!\not\!{k}+m_c)\gamma^\nu(\!\not\!{k}+m_c)\, ,
\end{eqnarray}
and  $t^n=\frac{\lambda^n}{2}$, the $\lambda^n$ is the Gell-Mann matrix
\cite{WangHuang3900,PRT85,Pascual-1984}.
We introduce  the $\langle\bar{q}_j\sigma_{\mu\nu}q_i \rangle$  comes from Fierz re-ordering  of the $\langle q_i \bar{q}_j\rangle$   to  absorb the gluons  emitted from other quark lines to  extract the mixed condensate   $\langle\bar{q}g_s\sigma G q\rangle$ \cite{WangHuang3900}.  Then we compute  all the Feynman diagrams  analytically, and finally obtain the QCD spectral densities through   dispersion relation,
\begin{eqnarray}\label{QCD-rho}
 \rho^1_{QCD}(s) &=&\frac{{\rm Im}\Pi^1_j(s)}{\pi}\, , \nonumber\\
\rho^0_{QCD}(s) &=&\frac{{\rm Im}\Pi^0_j(s)}{\pi}\, ,
\end{eqnarray}
where $j=\frac{1}{2}$, $\frac{3}{2}$, $\frac{5}{2}$.
We calculate the vacuum condensates up to dimension 13 which are vacuum expectations of the quark-gluon operators of the orders  $\mathcal{O}( \alpha_s^{k})$ with $k\leq 1$ consistently  \cite{WZG-penta-IJMPA,WangZG-Review,WangZG-Pcs4459-333,WangZG-XinQ-Pcs}.

Now we  match the hadron side with the QCD side of the correlation functions, take the quark-hadron duality below the continuum thresholds, and  obtain  two  QCD sum rules:
\begin{eqnarray}\label{QCDSR}
2M_{-}\lambda_{-}^2\exp\left( -\frac{M_{-}^2}{T^2}\right)&=& \int_{4m_c^2}^{s_0}ds \,\left[\sqrt{s}\rho_{QCD}^1(s)+\rho_{QCD}^{0}(s)\right]\,\exp\left( -\frac{s}{T^2}\right)\,  ,
\end{eqnarray}
\begin{eqnarray}\label{QCDSR-Positive}
2M_{+}\lambda_{+}^2\exp\left( -\frac{M_{+}^2}{T^2}\right)&=& \int_{4m_c^2}^{s^\prime_0}ds \,\left[\sqrt{s}\rho_{QCD}^1(s)-\rho_{QCD}^{0}(s)\right]\,\exp\left( -\frac{s}{T^2}\right)\,  .
\end{eqnarray}

If we neglect the hadronic couplings to the hidden-charm pentaquark states with positive parity, we could obtain two traditional QCD sum rules,
\begin{eqnarray}\label{Traditional-QCDSR-1}
\lambda_{-}^2\exp\left( -\frac{M_{-}^2}{T^2}\right)&=&\int_{4m_c^2}^{s_0}ds \,\rho^1_{QCD}(s)\exp\left( -\frac{s}{T^2}\right) \, ,
\end{eqnarray}
\begin{eqnarray}\label{Traditional-QCDSR-0}
M_{-}\lambda_{-}^2\exp\left( -\frac{M_{-}^2}{T^2}\right)&=&\int_{4m_c^2}^{s_0}ds \,\rho^0_{QCD}(s)\exp\left( -\frac{s}{T^2}\right) \, ,
\end{eqnarray}
with respect to the components $\Pi_j^1(p^2)$ and $\Pi^0_j(p^2)$ with the spins $j=\frac{1}{2}$, $\frac{3}{2}$, $\frac{5}{2}$, respectively.
However, such an approximation leads to contaminations as the hadronic couplings  are not zero.

According to the discussions in Ref.\cite{WangZG-XinQ-Pcs}, we define a parameter CTM to measure contaminations from the hidden-charm pentaquark states with positive parity,
\begin{eqnarray}
{\rm CTM}&=&\frac{\int_{4m_c^2}^{s_0}ds \,\left[\sqrt{s}\rho_{QCD}^1(s)-\rho_{QCD}^{0}(s)\right]\,\exp\left( -\frac{s}{T^2}\right)}{\int_{4m_c^2}^{s_0}ds \,\left[\sqrt{s}\rho_{QCD}^1(s)+\rho_{QCD}^{0}(s)\right]\,\exp\left( -\frac{s}{T^2}\right)}\, ,
\end{eqnarray}
by setting $s^\prime_0=s_0$ if the traditional QCD sum rules in Eqs.\eqref{Traditional-QCDSR-1}-\eqref{Traditional-QCDSR-0} are adopted. Direct calculations indicate that in the Borel windows,
  \begin{eqnarray}
{\rm CTM} &\sim & 10\%\,  \, {\rm or} \,\, 20\%\, ,
\end{eqnarray}
which are rather large and impair the predictive ability, the traditional  QCD sum rules in Eqs.\eqref{Traditional-QCDSR-1}-\eqref{Traditional-QCDSR-0} are discarded.

We differentiate   Eq.\eqref{QCDSR} with respect  to  $\frac{1}{T^2}$, then eliminate the pole residues $\lambda_{-}$ and acquire  the QCD sum rules for
 the pentaquark masses,
 \begin{eqnarray}
 M^2_{-} &=& \frac{-\int_{4m_c^2}^{s_0}ds \frac{d}{d(1/T^2)}\, \left[\sqrt{s}\rho_{QCD}^1(s)+\rho_{QCD}^{0}(s)\right]\,\exp\left( -\frac{s}{T^2}\right)}{\int_{4m_c^2}^{s_0}ds \, \left[\sqrt{s}\rho_{QCD}^1(s)+\rho_{QCD}^{0}(s)\right]\,\exp\left( -\frac{s}{T^2}\right)}\,  .
\end{eqnarray}

\section{Numerical results and discussions}
At the initial  points, we adopt  the standard values of the  vacuum condensates
$\langle\bar{q}q \rangle=-(0.24\pm 0.01\, \rm{GeV})^3$,
 $\langle\bar{q}g_s\sigma G q \rangle=m_0^2\langle \bar{q}q \rangle$,
$m_0^2=(0.8 \pm 0.1)\,\rm{GeV}^2$, $\langle \frac{\alpha_s
GG}{\pi}\rangle=0.012\pm0.004\,\rm{GeV}^4$    at the energy scale  $\mu=1\, \rm{GeV}$
\cite{SVZ79-1,SVZ79-2,PRT85,ColangeloReview}, and  take the $\overline{MS}$ mass $m_{c}(m_c)=(1.275\pm0.025)\,\rm{GeV}$
  from the Particle Data Group \cite{PDG}.
Moreover,  we take account of
the energy-scale dependence of  all the input parameters \cite{Narison-mix},
 \begin{eqnarray}
 \langle\bar{q}q \rangle(\mu)&=&\langle\bar{q}q\rangle({\rm 1 GeV})\left[\frac{\alpha_{s}({\rm 1 GeV})}{\alpha_{s}(\mu)}\right]^{\frac{12}{33-2n_f}}\, , \nonumber\\
  \langle\bar{q}g_s \sigma Gq \rangle(\mu)&=&\langle\bar{q}g_s \sigma Gq \rangle({\rm 1 GeV})\left[\frac{\alpha_{s}({\rm 1 GeV})}{\alpha_{s}(\mu)}\right]^{\frac{2}{33-2n_f}}\, ,\nonumber\\
  m_c(\mu)&=&m_c(m_c)\left[\frac{\alpha_{s}(\mu)}{\alpha_{s}(m_c)}\right]^{\frac{12}{33-2n_f}} \, ,\nonumber\\
\alpha_s(\mu)&=&\frac{1}{b_0t}\left[1-\frac{b_1}{b_0^2}\frac{\log t}{t} +\frac{b_1^2(\log^2{t}-\log{t}-1)+b_0b_2}{b_0^4t^2}\right]\, ,
\end{eqnarray}
  where $t=\log \frac{\mu^2}{\Lambda^2}$, $b_0=\frac{33-2n_f}{12\pi}$, $b_1=\frac{153-19n_f}{24\pi^2}$, $b_2=\frac{2857-\frac{5033}{9}n_f+\frac{325}{27}n_f^2}{128\pi^3}$,  $\Lambda_{QCD}=210\,\rm{MeV}$, $292\,\rm{MeV}$  and  $332\,\rm{MeV}$ for the flavors  $n_f=5$, $4$ and $3$, respectively  \cite{PDG}.
In this work, we study  the diquark-diquark-antiquark type $uudc\bar{c}$  pentaquark states,  it is better to choose the flavor numbers $n_f=4$, and evolve all the input parameters to a typical energy scale $\mu$, which satisfies  the energy scale formula,
\begin{eqnarray}\label{formula}
 \mu &=&\sqrt{M^2_{P}-(2{\mathbb{M}}_c)^2} \, ,
\end{eqnarray}
 with the updated value ${\mathbb{M}}_c=1.82\,\rm{GeV}$ \cite{Wang1508-EPJC,WangHuang-EPJC-1508-12,WangZG-EPJC-1509-12,
WangZG-NPB-1512-32,WZG-tetraquark-Mc}.

In the QCD sum rules for  the  baryon and pentaquark states with at least  one heavy quark,  we usually choose the continuum threshold parameters as $\sqrt{s_0}=M_{\rm gr}+ (0.5-0.8)\,\rm{GeV}$  \cite{Wang1508-EPJC,WZG-penta-IJMPA,WangHuang-EPJC-1508-12,
 WangZG-EPJC-1509-12,WangZG-NPB-1512-32,WangZhang-APPB,
 WangZG-Pcs4459-333,WangZG-XinQ-Pcs,Wang-cc-baryon-penta},  where  the subscript $\rm gr$ stands for  the ground states. As the ground state masses are unknown, the relation $\sqrt{s_0}=M_{\rm gr}+ (0.5-0.8)\,\rm{GeV}$  serves as a constraint in calculations.

 We acquire the  Borel  windows and continuum threshold parameters  via trial  and error, which are shown in Table \ref{Borel}. In the Borel windows, the pole contributions are about $(40-60)\%$, which is large enough to extract the pentaquark masses reliably in a systematic way. In fact, they are the largest pole contributions  in the QCD sum rules for the pentaquark states up to now.
 The pole contributions are defined by,
\begin{eqnarray}
{\rm{pole}}&=&\frac{\int_{4m_{c}^{2}}^{s_{0}}ds\,\rho_{QCD}\left(s\right)\exp\left(-\frac{s}{T^{2}}\right)} {\int_{4m_{c}^{2}}^{\infty}ds\,\rho_{QCD}\left(s\right)\exp\left(-\frac{s}{T^{2}}\right)}\, ,
\end{eqnarray}
 with the spectral densities  $\rho_{QCD}=\sqrt{s}\rho_{QCD}^1(s)+\rho_{QCD}^{0}(s)$.
  If we have not adopted  the energy scale formula in Eq.\eqref{formula}, we only obtain poor pole contributions, as the energy scale formula can enhance the pole contributions significantly and improve the convergent behavior of the operator product expansion significantly.

\begin{table}
\begin{center}
\begin{tabular}{|c|c|c|c|c|c|c|c|}\hline\hline
                  &$T^2(\rm{GeV}^2)$     &$\sqrt{s_0}(\rm{GeV})$    &$\mu(\rm{GeV})$  &pole          &$D(13)$         \\ \hline

$J^1(x)$          &$3.1-3.5$             &$4.96\pm0.10$             &$2.3$            &$(41-62)\%$   &$<1\%$      \\ \hline

$J^2(x)$          &$3.2-3.6$             &$5.10\pm0.10$             &$2.6$            &$(42-63)\%$   &$<1\%$      \\ \hline

$J^3(x)$          &$2.8-3.2$             &$4.85\pm0.10$             &$2.1$            &$(40-63)\%$   &$\sim 3\%$      \\ \hline

$J^4(x)$          &$3.1-3.5$             &$4.92\pm0.10$             &$2.2$            &$(40-62)\%$   &$\ll1\%$     \\ \hline

$J^1_\mu(x)$      &$3.3-3.7$             &$5.12\pm0.10$             &$2.6$            &$(41-62)\%$   &$< 1\%$     \\ \hline

$J^2_\mu(x)$      &$3.3-3.7$             &$5.02\pm0.10$             &$2.4$            &$(40-61)\%$   &$\ll 1\%$     \\ \hline

$J^3_\mu(x)$      &$3.2-3.6$             &$5.02\pm0.10$             &$2.4$            &$(41-62)\%$   &$< 1\%$     \\ \hline

$J^4_\mu(x)$      &$3.2-3.6$             &$5.02\pm0.10$             &$2.4$            &$(40-62)\%$   &$<1\%$     \\ \hline

$J^1_{\mu\nu}(x)$ &$3.3-3.7$             &$5.12\pm0.10$             &$2.6$            &$(41-62)\%$   &$< 1\%$     \\ \hline

$J^2_{\mu\nu}(x)$ &$3.3-3.7$             &$5.05\pm0.10$             &$2.4$            &$(40-60)\%$   &$< 1\%$     \\ \hline\hline
\end{tabular}
\end{center}
\caption{ The Borel  windows, continuum threshold parameters, optimal energy scales, pole contributions,   contributions of the vacuum condensates $D(13)$ for the currents with the isospin $I=\frac{1}{2}$. }\label{Borel}
\end{table}

\begin{figure}
\centering
\includegraphics[totalheight=6cm,width=7cm]{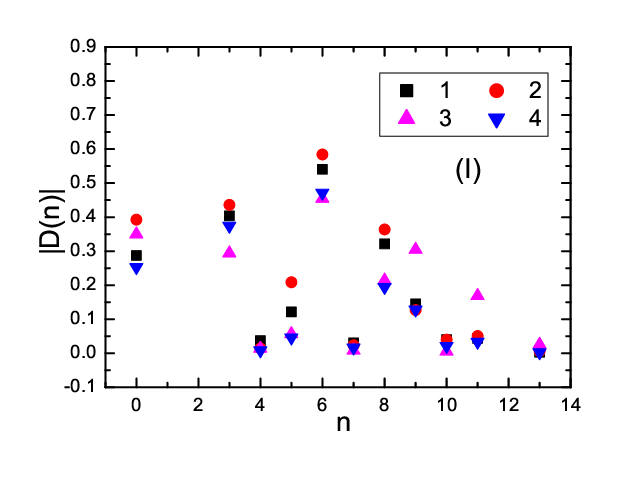}
\includegraphics[totalheight=6cm,width=7cm]{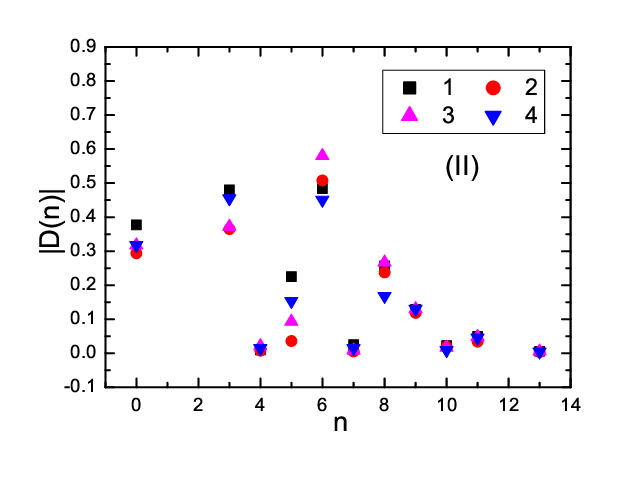}
\includegraphics[totalheight=6cm,width=7cm]{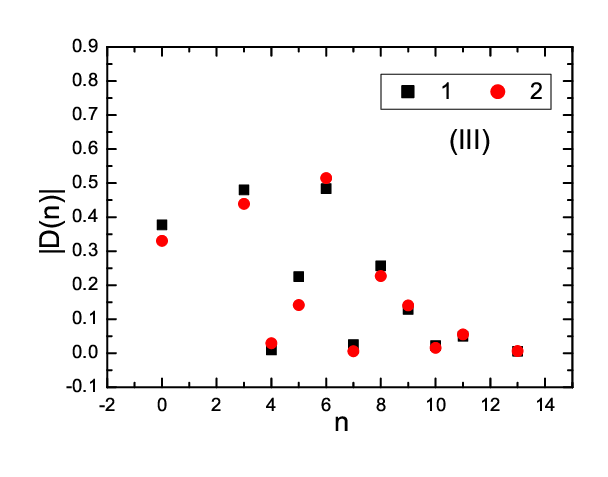}
 \caption{ The $|D(n)|$ with variations of the $n$ for the central values of the relevant parameters, where the (I), (II) and (III) denote  the spin $J=\frac{1}{2}$, $\frac{3}{2}$ and $\frac{5}{2}$, the $1$, $2$, $3$ and $4$ denote the series numbers of the currents. }\label{OPE-fig}
\end{figure}

In Fig.\ref{OPE-fig}, we plot the absolute values of the contributions of the vacuum condensates with the dimension $n$ for the central values of all the other  parameters, where the $D(n)$ are defined by,
   \begin{eqnarray}
D(n)&=&\frac{\int_{4m_{c}^{2}}^{s_{0}}ds\,\rho_{QCD,n}(s)\exp\left(-\frac{s}{T^{2}}\right)}
{\int_{4m_{c}^{2}}^{s_{0}}ds\,\rho_{QCD}\left(s\right)\exp\left(-\frac{s}{T^{2}}\right)}\, .
\end{eqnarray}
From the figure, we can see explicitly  that the $D(4)$ and $D(7)$ play a tiny role, the $D(0)$ and $D(3)$ play an important role and have the relation,
\begin{eqnarray}
D(3)> D(0)\gg |D(5)|\, ,
\end{eqnarray}
the contributions $D(5)$ are sizable.
No definite conclusion can be obtained if we only consider the contributions $D(n=0,3,4,5,7)$.
The $D(6)$ plays  a most  important role, and serves as a milestone to judge the convergent behavior of the operator product expansion. The $|D(n)|$ with $n\geq 6$ have the hierarchies,
\begin{eqnarray}
&&D(6)\gg |D(8)| \gg D(9) \gg D(10)\sim |D(11)| \gg D(13) \, , \nonumber\\
&&D(6)\gg |D(8)| \gg D(9) \gg |D(11)|> D(10) \gg D(13) \, , \nonumber\\
&&D(6)\gg  D(9)\gg |D(8)| \gg |D(11)| \gg D(13) \, , \nonumber\\
&&D(6)\gg |D(8)| \gg D(9) \gg |D(11)|> D(10) \gg D(13) \, ,
\end{eqnarray}
for the currents $J^1(x)$, $J^2(x)$, $J^3(x)$, $J^4(x)$, respectively,
\begin{eqnarray}
&&D(6)\gg |D(8)| \gg D(9) \gg |D(11)|> D(10) \gg D(13) \, , \nonumber\\
&&D(6)\gg |D(8)| \gg D(9) \gg |D(11)|> D(10) \gg D(13) \, , \nonumber\\
&&D(6)\gg |D(8)| \gg D(9) \gg |D(11)|> D(10) \gg D(13) \, , \nonumber\\
&&D(6)\gg |D(8)| \gg D(9) \gg |D(11)| \gg D(13) \, ,
\end{eqnarray}
for the currents $J_{\mu}^1(x)$, $J_{\mu}^2(x)$, $J_{\mu}^3(x)$, $J_{\mu}^4(x)$, respectively,
\begin{eqnarray}
&&D(6)\gg |D(8)| \gg D(9) \gg |D(11)|> D(10) \gg D(13) \, , \nonumber\\
&&D(6)\gg |D(8)| \gg D(9) \gg |D(11)|> D(10) \gg D(13) \, ,
\end{eqnarray}
for the currents $J_{\mu\nu}^1(x)$, $J_{\mu\nu}^2(x)$,  respectively, where we have  neglected  tiny values of the $D(7)$ and $D(10)$. The values of the $D(13)$  are shown explicitly in Table \ref{Borel}, i.e. $D(13)<1\%$ or $\ll 1\%$, except that $D(13)\sim 3\%$ in the case of the $J^3(x)$. All in all, the operator product expansion is convergent.

\begin{table}
\begin{center}
\begin{tabular}{|c|c|c|c|c|c|c|c|c|}\hline\hline
$[qq][qc]\bar{c}$ ($S_L$, $S_H$, $J_{LH}$, $J$) &$M(\rm{GeV})$   &$\lambda(10^{-3}\rm{GeV}^6)$ &Assignments        \\ \hline

$[ud][uc]\bar{c}$ ($0$, $0$, $0$, $\frac{1}{2}$)                      &$4.31\pm0.11$   &$1.40\pm0.23$                &$?\,P_c(4312)$         \\

$[ud][uc]\bar{c}$ ($0$, $1$, $1$, $\frac{1}{2}$)                      &$4.45\pm0.11$   &$3.02\pm0.48$                &$?\,P_c(4440/4457)$    \\

$[uu][dc]\bar{c}-[ud][uc]\bar{c}$ ($1$, $1$, $0$, $\frac{1}{2}$)     &$4.20\pm0.11$   &$2.24\pm0.40$                &         \\

$[uu][dc]\bar{c}-[ud][uc]\bar{c}$ ($1$, $0$, $1$, $\frac{1}{2}$)     &$4.25\pm0.11$   &$2.78\pm0.47$                &    \\

$[ud][uc]\bar{c}$ ($0$, $1$, $1$, $\frac{3}{2}$)                      &$4.45\pm0.11$   &$1.70\pm0.27$                &$?\,P_c(4440/4457)$   \\

$[uu][dc]\bar{c}-[ud][uc]\bar{c}$ ($1$, $0$, $1$, $\frac{3}{2}$)     &$4.34\pm0.11$   &$1.83\pm0.30$                &$?\,P_c(4312/4337)$  \\

$[uu][dc]\bar{c}-[ud][uc]\bar{c}$ ($1$, $1$, $2$, $\frac{3}{2}$)${}_3$     &$4.35\pm0.11$   &$3.10\pm0.51$                &$?\,P_c(4337/4380)$   \\

$[uu][dc]\bar{c}-[ud][uc]\bar{c}$ ($1$, $1$, $2$, $\frac{3}{2}$)${}_4$      &$4.34\pm0.11$   &$3.07\pm0.51$                &$?\,P_c(4312/4337)$   \\

$[ud][uc]\bar{c}$ ($0$, $1$, $1$, $\frac{5}{2}$)                      &$4.45\pm0.11$   &$1.70\pm0.27$                &$?\,P_c(4440/4457)$   \\

$[uu][dc]\bar{c}-[ud][uc]\bar{c}$ ($1$, $1$, $2$, $\frac{5}{2}$)     &$4.38\pm0.11$   &$1.76\pm0.29$                &$?\,P_c(4380)$       \\
\hline\hline
\end{tabular}
\end{center}
\caption{ The masses,  pole residues and possible assignments  of the hidden-charm pentaquark states with the isospin $I=\frac{1}{2}$ . }\label{mass}
\end{table}

\begin{figure}
\centering
\includegraphics[totalheight=6cm,width=7cm]{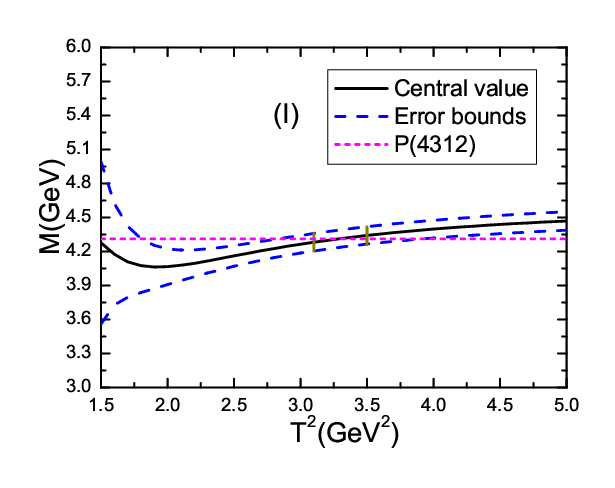}
\includegraphics[totalheight=6cm,width=7cm]{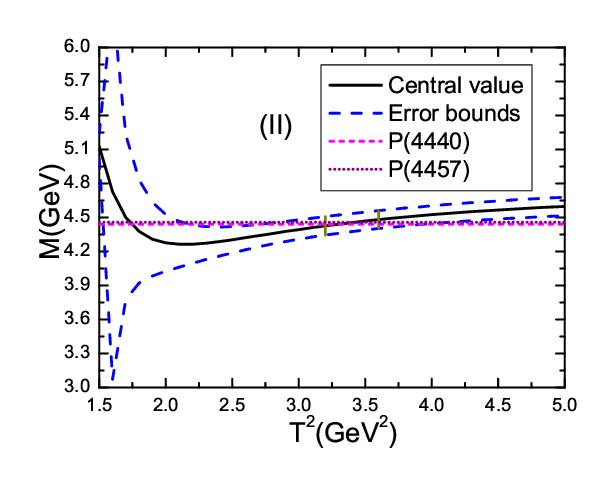}
\includegraphics[totalheight=6cm,width=7cm]{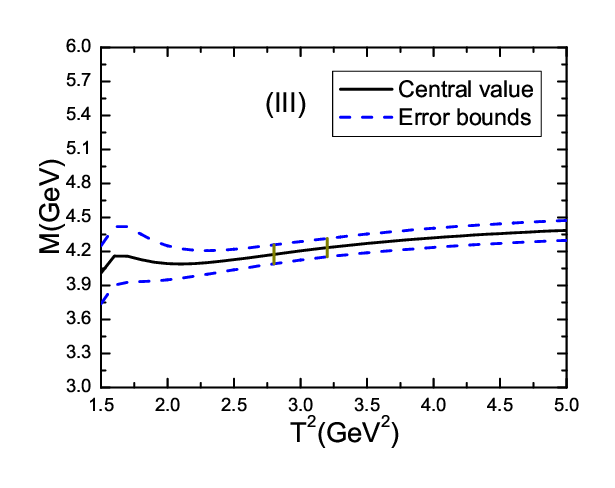}
\includegraphics[totalheight=6cm,width=7cm]{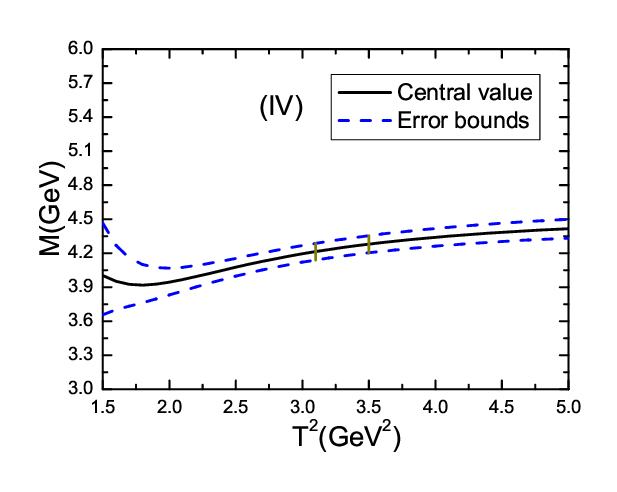}
  \caption{ The masses  with variations of the  Borel parameters $T^2$ for  the hidden-charm pentaquark states, where the (I), (II), (III)  and (IV)  denote the
   $[ud][uc]\bar{c}$ ($0$, $0$, $0$, $\frac{1}{2}$),
$[ud][uc]\bar{c}$ ($0$, $1$, $1$, $\frac{1}{2}$),
$[uu][dc]\bar{c}-[ud][uc]\bar{c}$ ($1$, $1$, $0$, $\frac{1}{2}$) and
$[uu][dc]\bar{c}-[ud][uc]\bar{c}$ ($1$, $0$, $1$, $\frac{1}{2}$)  pentaquark states, respectively. }\label{mass-12-fig}
\end{figure}

\begin{figure}
\centering
\includegraphics[totalheight=6cm,width=7cm]{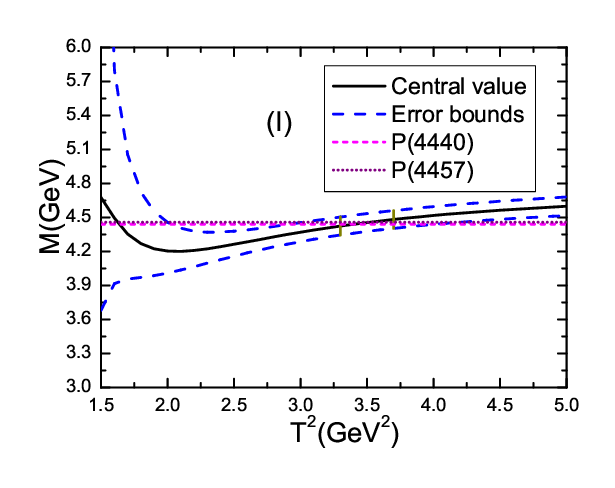}
\includegraphics[totalheight=6cm,width=7cm]{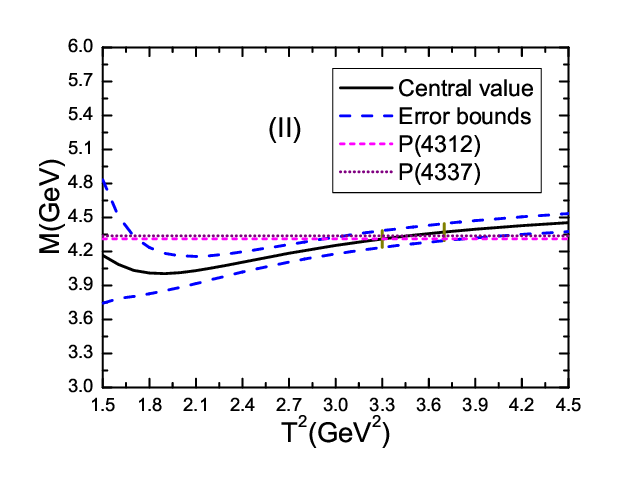}
\includegraphics[totalheight=6cm,width=7cm]{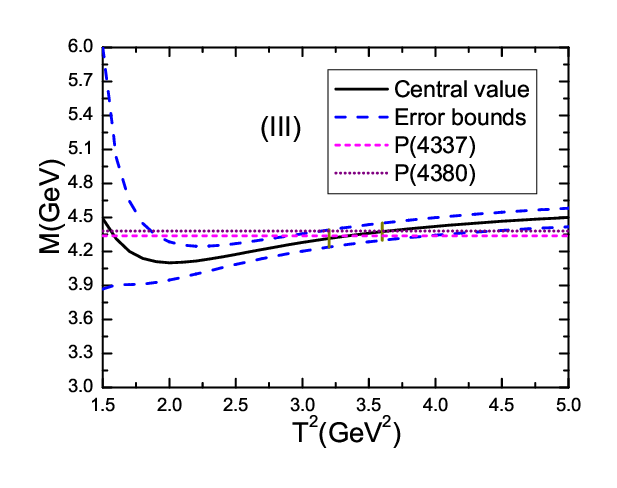}
\includegraphics[totalheight=6cm,width=7cm]{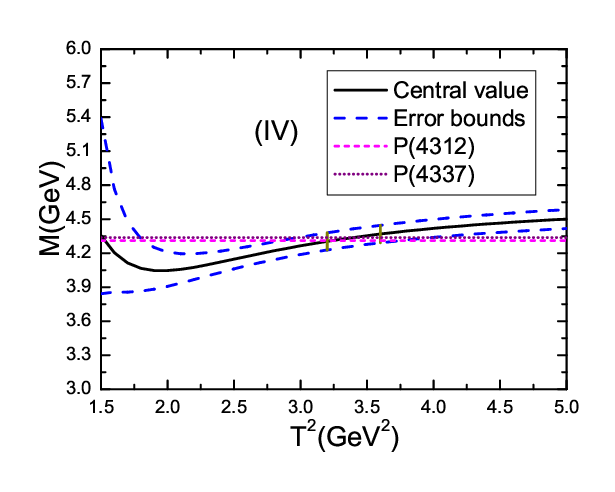}
  \caption{ The masses  with variations of the  Borel parameters $T^2$ for  the hidden-charm pentaquark states, where the (I) and (II), (III)  and (IV)  denote the $[ud][uc]\bar{c}$ ($0$, $1$, $1$, $\frac{3}{2}$),  $[uu][dc]\bar{c}-[ud][uc]\bar{c}$ ($1$, $0$, $1$, $\frac{3}{2}$),
$[uu][dc]\bar{c}-[ud][uc]\bar{c}$ ($1$, $1$, $2$, $\frac{3}{2}$)${}_{3}$ and
$[uu][dc]\bar{c}-[ud][uc]\bar{c}$ ($1$, $1$, $2$, $\frac{3}{2}$)${}_{4}$
    pentaquark states, respectively. }\label{mass-32-fig}
\end{figure}

\begin{figure}
\centering
\includegraphics[totalheight=6cm,width=7cm]{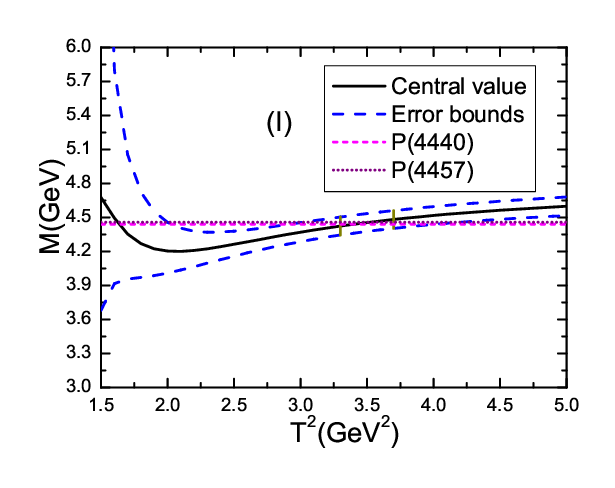}
\includegraphics[totalheight=6cm,width=7cm]{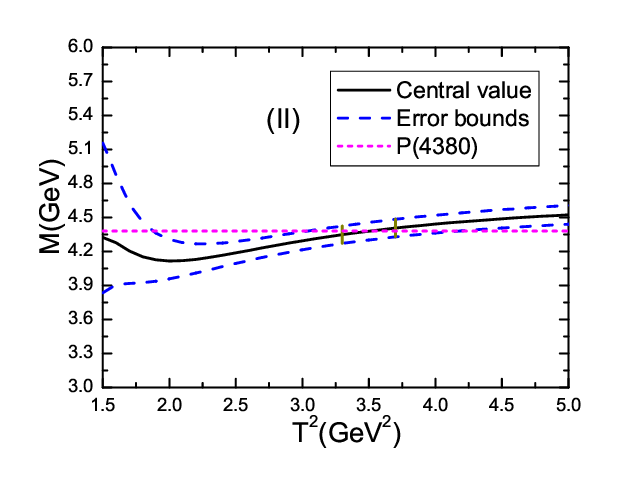}
  \caption{ The masses  with variations of the  Borel parameters $T^2$ for  the hidden-charm pentaquark states, where the (I) and (II) denote the
   $[ud][uc]\bar{c}$ ($0$, $1$, $1$, $\frac{5}{2}$) and $[uu][dc]\bar{c}-[ud][uc]\bar{c}$ ($1$, $1$, $2$, $\frac{5}{2}$)   pentaquark states, respectively. }\label{mass-52-fig}
\end{figure}

Then we take  account of all uncertainties  of the relevant  parameters,
and acquire   the masses and pole residues of
 the diquark-diquark-antiquark type  $uudc\bar{c}$ pentaquark states  with the isospin $I=\frac{1}{2}$, which are shown explicitly in Table \ref{mass} and Figs.\ref{mass-12-fig}-\ref{mass-52-fig}.

 From Tables \ref{Borel}-\ref{mass}, we can see explicitly that the central values have the relation $\sqrt{s_0}=M_{P}+(0.65-0.68)\,\rm{GeV}$, it is reasonable and it is expected to exclude contaminations from the higher resonances and continuum states. On the other hand, the energy scale formula,
 $\mu =\sqrt{M^2_{P}-(2{\mathbb{M}}_c)^2}$,  is satisfied very well, which warrants larger pole contributions and better convergent behaviors of the operator product expansion.

 In Figs.\ref{mass-12-fig}-\ref{mass-52-fig}, we plot the predicted pentaquark masses with variations of the Borel parameters at much larger ranges  than the Borel windows, where the Borel windows are indicated by the two short vertical lines. If we take $T_{max}^2$ and $T^2_{min}$ to represent the maximums and minimums of the Borel windows, at the region $T^2< T^2_{min}$, the flatness of the platforms becomes worse, which corresponds to the worse convergent behaviors of the operator product expansion; at the region $T^2> T^2_{max}$, the flatness of the platforms becomes better, which corresponds to the better convergent behaviors of the operator product expansion, however, the pole contributions becomes smaller, i.e. less than $40\%$. A compromise is obtained by choosing $T^2=T_{min}^2\sim T^2_{max}$.

 In Figs.\ref{mass-12-fig}-\ref{mass-52-fig}, we also present the experimental values of the masses of the $P_c(4312)$, $P_c(4337)$, $P_c(4380)$, $P_c(4440)$ and $P_c(4457)$ from the LHCb collaboration \cite{LHCb-4380,LHCb-Pc4312,LHCb-Pc4337}. Thus, we can see whether or not the assignments are robust  intuitively.

In Table \ref{mass}, we also present the possible assignments based on the predicted pentaquark masses, and we would like to give some interpretations in the following.

The predicted mass $4.31\pm0.11\,\rm{GeV}$ for the $[ud][uc]\bar{c}$ ($0$, $0$, $0$, $\frac{1}{2}$)  pentaquark state is in very good agreement with the experimental data $4311.9\pm0.7^{+6.8}_{-0.6} \mbox{ MeV}$ from the LHCb collaboration \cite{LHCb-Pc4312}, and
supports assigning the $P_c(4312)$ as the $[ud][uc]\bar{c}$ ($0$, $0$, $0$, $\frac{1}{2}$)  pentaquark state with the isospin-spin-parity $IJ^P=\frac{1}{2}{\frac{1}{2}}^{-}$. According to the arguments around Eq.\eqref{Jmu-5}, we cannot exclude assigning the $P_c(4312)$ as the $[ud][uc]\bar{c}$ ($0$, $0$, $0$, $\frac{3}{2}$)  pentaquark state with the isospin-spin-parity $IJ^P=\frac{1}{2}{\frac{3}{2}}^{-}$.

The predicted masses $4.45\pm0.11\,\rm{GeV}$ for the $[ud][uc]\bar{c}$ ($0$, $1$, $1$, $\frac{1}{2}$), $[ud][uc]\bar{c}$ ($0$, $1$, $1$, $\frac{3}{2}$) and  $[ud][uc]\bar{c}$ ($0$, $1$, $1$, $\frac{5}{2}$) pentaquark states are all in very good agreement with the experimental data $4440.3\pm1.3^{+4.1}_{-4.7} \mbox{ MeV}$ and
$4457.3\pm0.6^{+4.1}_{-1.7} \mbox{ MeV}$ from the LHCb collaboration \cite{LHCb-Pc4312}, and
supports assigning the $P_c(4440)$ and $P_c(4457)$ as the  $[ud][uc]\bar{c}$ ($0$, $1$, $1$, $\frac{1}{2}$),  $[ud][uc]\bar{c}$ ($0$, $1$, $1$, $\frac{3}{2}$) and  $[ud][uc]\bar{c}$ ($0$, $1$, $1$, $\frac{5}{2}$) pentaquark states with the isospin-spin-parity $IJ^P=\frac{1}{2}{\frac{1}{2}}^{-}$, $\frac{1}{2}{\frac{3}{2}}^{-}$ and $\frac{1}{2}{\frac{5}{2}}^{-}$, respectively.                                At the present time, it is very difficult to assign the $P_c(4440)$ and $P_c(4457)$ unambiguously even the strong decays are studied theoretically, as they have analogous narrow widths with large uncertainties.

The predicted masses $4.34\pm0.11\,\rm{GeV}$,  $4.35\pm0.11\,\rm{GeV}$ and  $4.34\pm0.11\,\rm{GeV}$ for the $[uu][dc]\bar{c}-[ud][uc]\bar{c}$ ($1$, $0$, $1$, $\frac{3}{2}$),  $[uu][dc]\bar{c}-[ud][uc]\bar{c}$ ($1$, $1$, $2$, $\frac{3}{2}$)${}_3$    and $[uu][dc]\bar{c}-[ud][uc]\bar{c}$ ($1$, $1$, $2$, $\frac{3}{2}$)${}_4$  pentaquark states respectively  are   in very good agreement with the experimental data $4311.9\pm0.7^{+6.8}_{-0.6} \mbox{ MeV}$ and $4337 \ ^{+7}_{-4} \ ^{+2}_{-2} \mbox{ MeV}$ from the LHCb collaboration \cite{LHCb-Pc4312,LHCb-Pc4337}, and support assigning the     $P_c(4312)$ and  $P_c(4337)$ as the $[uu][dc]\bar{c}-[ud][uc]\bar{c}$ ($1$, $0$, $1$, $\frac{3}{2}$),  $[uu][dc]\bar{c}-[ud][uc]\bar{c}$ ($1$, $1$, $2$, $\frac{3}{2}$)${}_3$    and $[uu][dc]\bar{c}-[ud][uc]\bar{c}$ ($1$, $1$, $2$, $\frac{3}{2}$)${}_4$  pentaquark states with the isospin-spin-parity $IJ^P=\frac{1}{2}{\frac{3}{2}}^{-}$.

The predicted mass $4.38\pm0.11\, \rm{GeV}$ for the $[uu][dc]\bar{c}-[ud][uc]\bar{c}$ ($1$, $1$, $2$, $\frac{5}{2}$) pentaquark state is in very good agreement with the experimental data  $4380\pm 8\pm 29 \mbox{ MeV}$ from the LHCb collaboration \cite{LHCb-4380}, and supports assigning the   $P_c(4380)$ as the $[uu][dc]\bar{c}-[ud][uc]\bar{c}$ ($1$, $1$, $2$, $\frac{5}{2}$) pentaquark state
with the isospin-spin-parity $IJ^P=\frac{1}{2}{\frac{5}{2}}^{-}$. Furthermore, the LHCb collaboration cannot exclude the assignment $J^P={\frac{5}{2}}^-$. The predicted mass $4.35\pm0.11\,\rm{GeV}$ for the $[uu][dc]\bar{c}-[ud][uc]\bar{c}$ ($1$, $1$, $2$, $\frac{3}{2}$)${}_3$  pentaquark state indicates that it is marginal to assign the $P_c(4380)$ as the $[uu][dc]\bar{c}-[ud][uc]\bar{c}$ ($1$, $1$, $2$, $\frac{3}{2}$)${}_3$  pentaquark state with the isospin-spin-parity $IJ^P=\frac{1}{2}{\frac{3}{2}}^{-}$. Without more experimental data and more theoretical works on the decays and productions, we cannot assign those $P_c$ states unambiguously.

A typical prediction is the mass (see the mass  $4.20\pm0.11\,\rm{GeV}$ of the third state in Table \ref{mass}) of the $[uu][dc]\bar{c}-[ud][uc]\bar{c}$ ($1$, $1$, $0$, $\frac{1}{2}$) pentaquark state with the isospin-spin-parity   $IJ^P=\frac{1}{2}{\frac{1}{2}}^{-}$, which is the lowest hidden-charm pentaquark state,  and (the central value of the mass) lies just above the $\bar{D}\Lambda_c$ threshold $4151\,\rm{MeV}$ from the Particle Data Group \cite{PDG}, and serves as a milestone for the mass spectrum of the hidden-charm pentaquark states with the isospin $I=\frac{1}{2}$.  We can search for those $P_c$ states in the two-body strong decays,
\begin{eqnarray}
P_{c}&\to& \bar{D}\Sigma_c\, , \, \bar{D}\Lambda_c\, , \, \bar{D}^*\Sigma_c\, , \, \bar{D}^*\Sigma_c^*\, , \, \bar{D}^*\Lambda_c\, , \, J/\psi p \, , \, \eta_c p \, ,
\end{eqnarray}
by precisely measuring the masses, widths, spins and parities, then diagnose their nature.

\section{Conclusion}
  In this work, we distinguish the isospin unambiguously to construct the   diquark-diquark-antiquark type local five-quark currents with the isospin $I=\frac{1}{2}$,  and study the $uudc\bar{c}$ pentaquark states with the QCD sum rules systematically. We compute  all the  vacuum condensates up to dimension $13$  consistently, which are vacuum expectations of the quark-gluon operators of the order $\mathcal{O}(\alpha_s^k)$ for $k\leq1$, distinguish the contributions of the hidden-charm pentaquark states with negative and positive parity unambiguously,  and adopt the energy scale formula $\mu=\sqrt{M_{P}^2-(2{\mathbb{M}}_c)^2}$ to choose  the optimal  energy scales of the QCD spectral densities. Then we obtain the mass spectrum of the  diquark-diquark-antiquark type $uudc\bar{c}$ pentaquark states with the isospin-spin-parity $IJ^P=\frac{1}{2}{\frac{1}{2}}^-$, $\frac{1}{2}{\frac{3}{2}}^-$ and $\frac{1}{2}{\frac{5}{2}}^-$, and make possible assignments of the $P_c(4312)$, $P_c(4337)$, $P_c(4380)$, $P_c(4440)$ and $P_c(4457)$.   Furthermore, we obtain the lowest hidden-charm pentaquark state which lies just above the $\bar{D}\Lambda_c$ threshold and can be confronted to the experimental data in the future.

 There have been several possible interpretations for those $P_c$ states, such as pentaquark states, (dynamically generated) molecular states,  triangle singularities, etc. All the interpretations have both advantages and shortcomings. In the present work, we make possible assignments based on the predicted masses from the QCD sum rules. Although all the $P_c$ states could find their suitable positions and the hidden-charm pentaquark scenario is robust, we cannot assign those $P_c$ states unambiguously with the masses alone, furthermore, the theoretically  predicted states with definite isospin are more than the experimentally observed ones. We should study the decays and productions  systematically to reach un-unambiguous assignments based on more precise experimental data.

\section*{Acknowledgements}
This  work is supported by National Natural Science Foundation, Grant Number
12575083.

\end{document}